\documentclass[%
reprint,
superscriptaddress,
 amsmath,amssymb,
 aps,
]{revtex4-1}

\usepackage{xcolor}
\usepackage{graphicx}
\usepackage{dcolumn}
\usepackage{bm}

\newcommand{\ket}[1]{|#1\rangle}
\newcommand{\braket}[2]{\langle #1|#2\rangle}
\begin{document}
\title{Basis independent tomography of complex vectorial light fields by Stokes projections}

\author{Adam Selyem}
\affiliation{School of Physics and Astronomy, University of Glasgow, Glasgow G12 8QQ, Scotland}
\author{Carmelo Rosales-Guzm\'an}
\altaffiliation[Now at ]{Wang Da-Heng Collaborative Innovation Center for Quantum manipulation \& Control, Harbin University of Science and Technology, Harbin 150080, China}
\affiliation{School of Physics, University of the Witwatersrand, Johannesburg 2050, South Africa}
\author{Sarah Croke}
\affiliation{School of Physics and Astronomy, University of Glasgow, Glasgow G12 8QQ, Scotland}
\author{Andrew Forbes}
\email{Andrew.Forbes@wits.ac.za}  
\affiliation{School of Physics, University of the Witwatersrand, Johannesburg 2050, South Africa}
\author{Sonja Franke-Arnold}
\email{Sonja.Franke-Arnold@glasgow.ac.uk} 
\affiliation{School of Physics and Astronomy, University of Glasgow, Glasgow G12 8QQ, Scotland}

\begin{abstract}
Complex vectorial light fields, non-separable in their polarization and spatial degree of freedom, are of relevance in a wide variety of fields encompassing microscopy, metrology, communication and topological studies.  Controversially, they have been suggested  as analogues to quantum entanglement, raising fundamental questions on the relation between non-separability in classical systems, and entanglement in quantum systems.  Here we propose and demonstrate basis-independent tomography of arbitrary vectorial light fields by relating their concurrence to spatially resolved Stokes projections. 
We generate vector fields with controllable non-separability using a novel compact interferometer that incorporates a digital micro-mirror device (DMD), thus offering a holistic toolbox for the generation and quantitative analysis of arbitrary vectorial light fields.     
\end{abstract}

\maketitle

 
The fact that light is a complex vector field rather than a scalar field affects many optical properties, including focusing,  propagation through inhomogeneous media,  interaction with matter and certain non-linear effects.  Over the last decade we have learned to design vector light fields with specified phase, amplitude and polarization profiles.  Contrary to homogeneously polarized scalar fields, vector light fields feature a non-homogeneous polarization distribution. The inherent correlations between the spatial and polarization degree of freedom mimic quantum entanglement - though not non-locality \cite{Spreeuw1998,Holleczek2011,Qian2015,Perez-Garcia2015,Aiello2015,Li2016}.  
These correlations allow tracking spatial location based on polarization measurements \cite{BergJohansen2015} and identifying polarization states from spatial camera measurements \cite{Radwell2018}, and have been used as a resource for a wide range of applications \cite{Rosales2018Review} including laser material processing \cite{Meier2007}, optical manipulations \cite{Zhan2009,Woerdemann2013}, high resolution microscopy \cite{Chen2013}, as well as classical and quantum communication and information processing \cite{Roadmap,Milione2015,DAmbrosio2016,Ndagano2017,Ndagano2018}.

Vector beams can be generated directly from a laser resonator \cite{Naidoo2016} or external to it by using geometrical phase elements \cite{Niv2004,Marrucci2006,Yachao2014,Radwell2016} or optical interferometers \cite{Tidwell1990}, while digital holography, be it in the form of spatial light modulators (SLMs) or digital micromirror devices (DMDs),  provides the most flexible and versatile approach \cite{Davis2000,Maurer2007,Moreno2012,Rong2014,Chen2014,Gong2014,Ren2015,Otte2016,Mitchell2016,Rosales2017,Mitchell2016,Mitchell2017,Liu2018}. 

Vector beams can be quantified and analyzed using a variety of approaches, including geometric phase measurement \cite{DErrico2017}, shear interferometry \cite{Khajavi2018}, Bell violations \cite{Borges2010,Qian2015,McLaren2015}, and quantum state tomographies via projective measurements \cite{Ndagano2016}. Despite these advances, all previously suggested measures of non-separability assess only a predefined subspace of vector beams, the surface of a specific higher-order Poincar\'{e} sphere (HOPS) \cite{Beckley2010,Bauer2015,Milione2011}.   
In case of projection measurements this subspace is further limited by testing a discretized selection of basis states.  Yet it is known that quantum entanglement persists independent of measurement basis, a fact that must translate to the classical non-separable vector states.  
 
Here we demonstrate that Stokes measurements are sufficient to fully characterize the degree of non-separability of complex light fields, as determined by its concurrence.  This allows us to characterize any vector field without \textit{a priori} knowledge of its spatial degree of freedom.  Our approach formally links concurrence to the well-known Stokes parameters of optical fields.  We confirm the validity of our basis-independent method by comparing it with the established method of basis-dependent state tomography. 

We have developed a compact, fast and inexpensive interferometric device that can create arbitrary complex vector light fields, exploiting the fact that DMDs can modulate any polarization state. The ability to generate light fields with varying degrees of correlation between spatial and polarization degrees of freedom is crucial for the concurrence measurements reported in this paper.  More generally, however, we expect that our device will be of importance in any application requiring full control over vector light fields, e.g. for super-resolution microscopy, sensing or fundamental studies of topological behavior.  Thus we offer a complete toolbox for both the creation and quantitative analysis of complex vectorial light fields.

\textbf{Theoretical concept.}  The formal analogy between non-separable vector beams and entangled quantum states allows the application of quantum concepts to classical light beams.   Concurrence, a quantum measure of entanglement in two dimensions, has been identified as a good measure of the degree of non-separability of vector beams, a vector quality factor \cite{Ndagano2016, McLaren2015}. 
Here we will define the concurrence in terms of Stokes projections, providing a spatially resolved and basis-independent measure of entanglement.

Any (paraxial) vectorial light beam can be written as  $ \vec{u}(\vec{r}_\perp)=u_h(\vec{r}_\perp) \hat{h}+u_v(\vec{r}_\perp) \hat{v}$, where $u_{h,v}(\vec{r}_\perp)$ are the complex spatial profiles of the horizontal and vertical polarization components and $\vec{r}_\perp=(x,y)$ is the transverse position.   To emphasize the formal analogy with entangled states we denote this field using bra-ket notation, 
\begin{equation} \label{Psi}
|\Psi\rangle =  |\widetilde{\psi}_{h}\rangle |h\rangle +|\widetilde{\psi}_{v}\rangle  |v\rangle,
\end{equation}
where $\{ |h\rangle, |v \rangle \}$  is the basis for the polarization degree of freedom, and $|\widetilde{\psi}_{h,v}\rangle$  are arbitrary (unnormalised) spatial states. Following \cite{Toppel2013}, we define position eigenstates in $x$ and $y$, such that, up to normalisation, $u_{h,v}(\vec{r}_\perp) = \braket{x,y}{\widetilde{\psi}_{h,v}}$. The states $\ket{\widetilde{\psi}_{h,v}}$ may then be expressed in the basis $\ket{x,y}$, and parameterized in terms of their amplitude and phase as 
\begin{equation} 
|\widetilde{\psi}_{h,v} \rangle = \int {\rm d}x {\rm d}y |\widetilde{\psi}_{h,v}(\vec{r}_\perp)|\exp[i\phi_{h,v}(\vec{r}_\perp)]|x,y\rangle.
\label{param}
\end{equation}
We assume the overall state to be pure, a reasonable assumption for most experimentally produced vector beams.

The state (\ref{Psi}) is associated with spatially varying local Stokes parameters 
$$
\vec{S}(\vec{r}_\perp)=\begin{pmatrix}   S_0(\vec{r}_\perp)\\S_1(\vec{r}_\perp)\\S_2(\vec{r}_\perp)\\S_3(\vec{r}_\perp)\end{pmatrix}= \begin{pmatrix} I_h(\vec{r}_\perp)+I_v(\vec{r}_\perp)\\I_h(\vec{r}_\perp)-I_v(\vec{r}_\perp)\\I_d(\vec{r}_\perp)-I_a(\vec{r}_\perp)\\I_r(\vec{r}_\perp)-I_l(\vec{r}_\perp)\end{pmatrix} .
$$
Here $I_{h,v,d,a,r,l}(\vec{r}_\perp)$ are the intensity profiles of the horizontal, vertical, diagonal, antidiagonal, right and left circular polarized light components respectively, given by $
I_h(\vec{r}_\perp) = I |\langle h| \langle x,y| \Psi \rangle|^2$, and $I$ is the total intensity in the vector beam.  

Local Stokes parameter measurements give us enough information to find the amplitudes $|\widetilde{\psi}_{h,v}(\vec{r}_\perp)|$,  and the relative phases $\phi_h(\vec{r}_\perp) - \phi_v(\vec{r}_\perp)$,  i.e. to perform a tomography of the vector beam apart from an unknown position dependent overall phase:
\begin{eqnarray} \label{Stokes}
S_1(\vec{r}_\perp) & =& I \left( |\widetilde{\psi}_{h}(\vec{r}_\perp)|^2 - |\widetilde{\psi}_{v}(\vec{r}_\perp)|^2 \right), \nonumber\\
S_2(\vec{r}_\perp) &=& 2 I |\widetilde{\psi}_h(\vec{r}_\perp)|   |\widetilde{\psi}_v(\vec{r}_\perp)|  \cos \left( \phi_h(\vec{r}_\perp) - \phi_v(\vec{r}_\perp) \right) , \nonumber\\
S_3(\vec{r}_\perp) &=& 2 I |\widetilde{\psi}_h(\vec{r}_\perp)|   |\widetilde{\psi}_v(\vec{r}_\perp)|  \sin \left( \phi_h(\vec{r}_\perp) - \phi_v(\vec{r}_\perp)\right).  
\end{eqnarray}
In the following we show that measurement of the global Stokes parameters is sufficient to quantify the degree of non-separability of a vector beam.  Using eqn (\ref{param}), the overlap between the spatial states is given by
$$
\langle\widetilde{\psi}_h|\widetilde{\psi}_v\rangle = \int {\rm d}\vec{r}_\perp |\widetilde{\psi}_h(\vec{r}_\perp)| |\widetilde{\psi}_v(\vec{r}_\perp)| e^{i \left( \phi_v(\vec{r}_\perp) - \phi_h(\vec{r}_\perp) \right)}.
$$
We will express our state (\ref{Psi}) in a form for which the concurrence may be readily calculated \cite{Wootters2001}: 
\begin{eqnarray} \label{Concurrence_general}
|\Psi\rangle = a  |\psi_{+}\rangle|h\rangle + b  |\psi_{-}\rangle|h\rangle + c |\psi_{+}\rangle|v\rangle + d |\psi_{-}\rangle|v\rangle, \end{eqnarray} 
for some orthonormal states $\ket{\psi_{+}}$, $\ket{\psi_{-}}$ of the spatial degree of freedom, with the concurrence given by $ C(|\Psi\rangle) = 2 |ad-bc|$ . We therefore need to identify a convenient orthonormal pair of spatial states.

In previous work \cite{Ndagano2016, McLaren2015,Aiello2015}, the subspace of the spatial degree of freedom containing the correlations with the polarization degree of freedom is  defined with respect to a {\emph predefined} orthonormal basis $\{\ket{\psi_{+}}$, $\ket{\psi_{-}} \}$. 
Here, we do not wish to assume anything about this subspace, and instead extract this information directly from our Stokes parameter measurements.
The subspace of interest is spanned by the states $\ket{\widetilde{\psi}_h}$, $\ket{\widetilde{\psi}_v}$, however these are not orthonormal in general, and we need to apply some orthogonalization procedure (e.g. Gram-Schmidt). We define normalized states $|\psi_{h,v}\rangle=p_{h,v}^{-1/2}|\widetilde{\psi}_{h,v}\rangle$ where 
$p_{h,v}=|\langle \widetilde{\psi}_{h,v}|\widetilde{\psi}_{h,v}\rangle|.$ For convenience and without loss of generality, we choose the global phase of $|\psi_v\rangle $ so that  $\langle\psi_h | \psi_v\rangle = \langle\psi_h | \psi_v\rangle ^\ast$, allowing us to define the orthonormal spatial states \begin{eqnarray*}
|\psi_{\pm}\rangle & = & N_{\pm} \left( | \psi_h\rangle \pm | \psi_v\rangle \right), 
\end{eqnarray*}
where we have defined normalization constants $N_{\pm}=(2(1\pm|\langle\psi_h | \psi_v\rangle| ))^{-1/2}.$

It is readily verified that our general vector state (\ref{Psi}) then takes the form
\begin{eqnarray*}
|\Psi\rangle &=&  \frac{\sqrt{p_h}}{2 N_+}   |\psi_{+}\rangle |h\rangle + \frac{\sqrt{p_h}}{2 N_-}   |\psi_{-}\rangle |h\rangle\\
&& +  \frac{\sqrt{p_v}}{2 N_+}   |\psi_{+}\rangle|v\rangle  - \frac{\sqrt{p_v}}{2 N_-}  |\psi_{-}\rangle|v\rangle .
\end{eqnarray*}
In comparison with (\ref{Concurrence_general}), we can now read off the concurrence as:
\begin{eqnarray} \label{concurrence_spatial}
\notag C(|\Psi\rangle)&=& 2 \sqrt{p_h p_v} \sqrt{1 - |\langle\psi_h |\psi_v\rangle|^2} \\ & =&
 2 \sqrt{\langle\widetilde{\psi}_h |\widetilde{\psi}_h\rangle \langle\widetilde{\psi }_v|\widetilde{\psi}_v\rangle- |\langle\widetilde{\psi}_h |\widetilde{\psi}_v\rangle|^2}.
\end{eqnarray}
Thus the concurrence depends on the relative amplitudes of the spatial modes, and their overlap.  For orthogonal modes $|\widetilde{\psi}_{h,v}\rangle$ the concurrence is maximal and decreases to zero as the overlap between the modes increases.  Note that for spatial modes shaped solely in their amplitude the overlap term can be deduced directly from intensity measurements.  For complex spatial modes the concurrence in (\ref{concurrence_spatial}) can instead be expressed in terms of the global Stokes parameters, i.e.~$\vec{S}=\int d\vec{r}_{\bot} \vec{S} (\vec{r}_{\bot})$, using the expressions in (\ref{Stokes}):
\begin{eqnarray} \label{Concurrence_Stokes}
C(|\Psi\rangle) = \sqrt{1 - \frac{S_1^2}{S_0^2} - \frac{S_2^2}{S_0^2} - \frac{S_3^2}{S_0^2}}.
\end{eqnarray}

While theoretically the first Stokes parameter is  $S_0=I=I_h + I_v = I_d + I_a =I_r + I_l ,$  the different optical properties of the polarization optics used to measure the intensities in the linear and circular polarization basis usually lead to small differences between the expressions.  To allow for this, we use the analogue formulation of the concurrence: 
\begin{equation} \label{Concurrence_Experiment}
C(|\Psi\rangle) = \sqrt{1 - \left(\frac{I_h - I_v}{I_h + I_v}\right)^2 - \left(\frac{I_d - I_a}{I_d + I_a}\right)^2 - \left(\frac{I_r - I_l}{I_r + I_l}\right)^2}.
\end{equation}

We finally return to the assumption throughout, that the overall state is a pure state. The concurrence for pure states [see (\ref{Concurrence_Stokes})] is a measure of the degree of polarization of the spatially averaged beam. This is a good measure of correlation if we can assume that all fluctuations in the polarization state of the beam are due to correlations with the spatial degree of freedom, which is a reasonable assumption for our state preparation technique.
Nevertheless, noise may be introduced in the measurement, primarily due to dark counts in the detectors. Further, the effect of noise is to \emph{overestimate} the concurrence, generating non-deliberate correlations due to intensity fluctuations.
A related phenomenon has been noted for quantum entanglement, where vacuum fluctuations generate an increase in negativity \cite{Chou2015}.
We will show in the next section that the removal of high-frequency modes allows us to suppress such noise in our data, supporting the assumption of an overall pure state.  The spatially resolved Stokes parameter measurements then allow us to quantify the non-separability of our vector states.

\begin{figure*}[htb]
\centering
\includegraphics[width=1.0\textwidth]{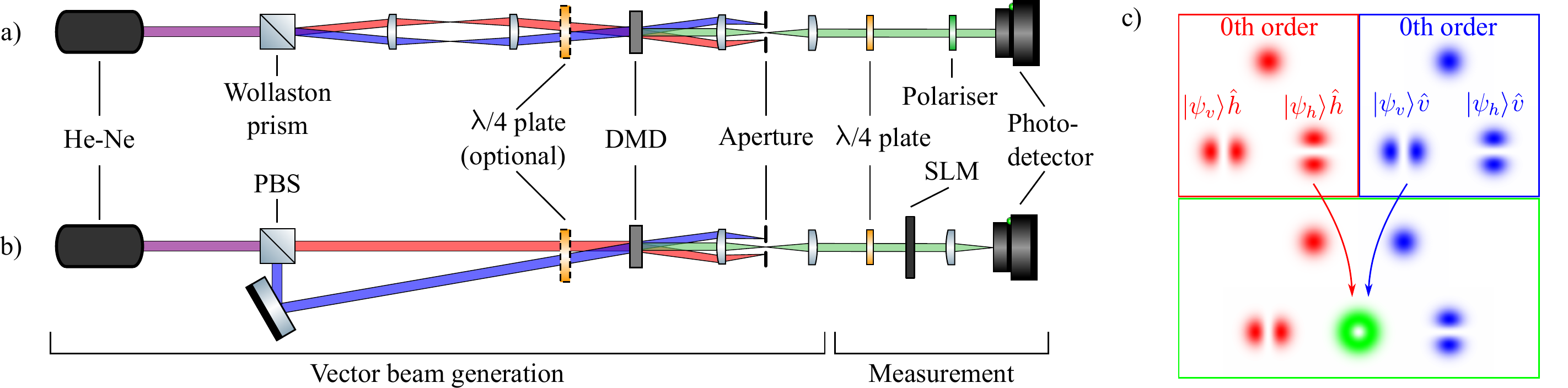}
\caption{Schematic apparatus for the generation of arbitrary vector beams using a DMD and a Wollaston prism (a) or a polarizing beam splitter (b), with the DMD shown in unfolded view. The complex fields of two orthogonally polarized beams (shown in red and blue) are coherently overlapped in the appropriate diffraction orders of the multiplexed grating displayed on the DMD to generate spatially varying polarization (green), illustrated in (c) for a radially polarized beam.  
The generated beams can be analyzed with spatially resolved Stokes measurements (a) or projection measurements using an SLM (b).
}
\label{setup}
\end{figure*}

\textbf{Experimental realization and results.} We test the concept of our basis independent concurrence measurement on arbitrary vector beams of the form $ \vec{u}(\vec{r}_\perp)=u_h(\vec{r}_\perp) \hat{h}+u_v(\vec{r}_\perp) \hat{v}.$
To this end we have developed a compact interferometric device that uses a DMD to shape the phase and amplitude of the vertical and horizontal polarization components of a desired light field independently, using techniques described e.g.~in \cite{Mirhosseini2013,Mitchell2016}.  The DMD acts, at the same time, effectively as the second beam splitter in a Mach-Zehnder type interferometer, as shown in Fig.~\ref{setup} a) and b). 

An expanded, collimated and diagonally polarized laser beam 
is separated into its vertical and horizontal polarization components by a polarizing beam splitter, e.g. a Wollaston prism.  Both polarization components are directed toward the DMD, where they spatially overlap but impinge under slightly different angles (separated by $\approx 1.5^\circ$).  The DMD displays a multiplexed hologram,  consisting of two holograms corresponding to the transformation of the initial amplitudes into the desired spatial wavefunctions $u_h(\vec{r}_\perp)$ and $u_v(\vec{r}_\perp)$, each superimposed with a different linear diffraction grating which separates the resulting profiles. 
The grating periods in combination with the different input angles are carefully chosen such that the appropriate diffraction orders of the horizontally and vertically polarized light fields overlap, thus generating the desired complex vector field as shown in Fig.~\ref{setup}c).  Any unwanted diffraction orders are removed by placing a spatial filter in the Fourier plane of a telescope which images the DMD plane. We note that the insertion of a quarter wave plate prior to the DMD generates the vector field encoded in circular polarization components, $ \vec{u}(\vec{r}_\perp)=u_l(\vec{r}_\perp) \hat{l}+u_r(\vec{r}_\perp) \hat{r}.$  

\begin{figure*}[bt]
\centering
\includegraphics[width=1.0\textwidth]{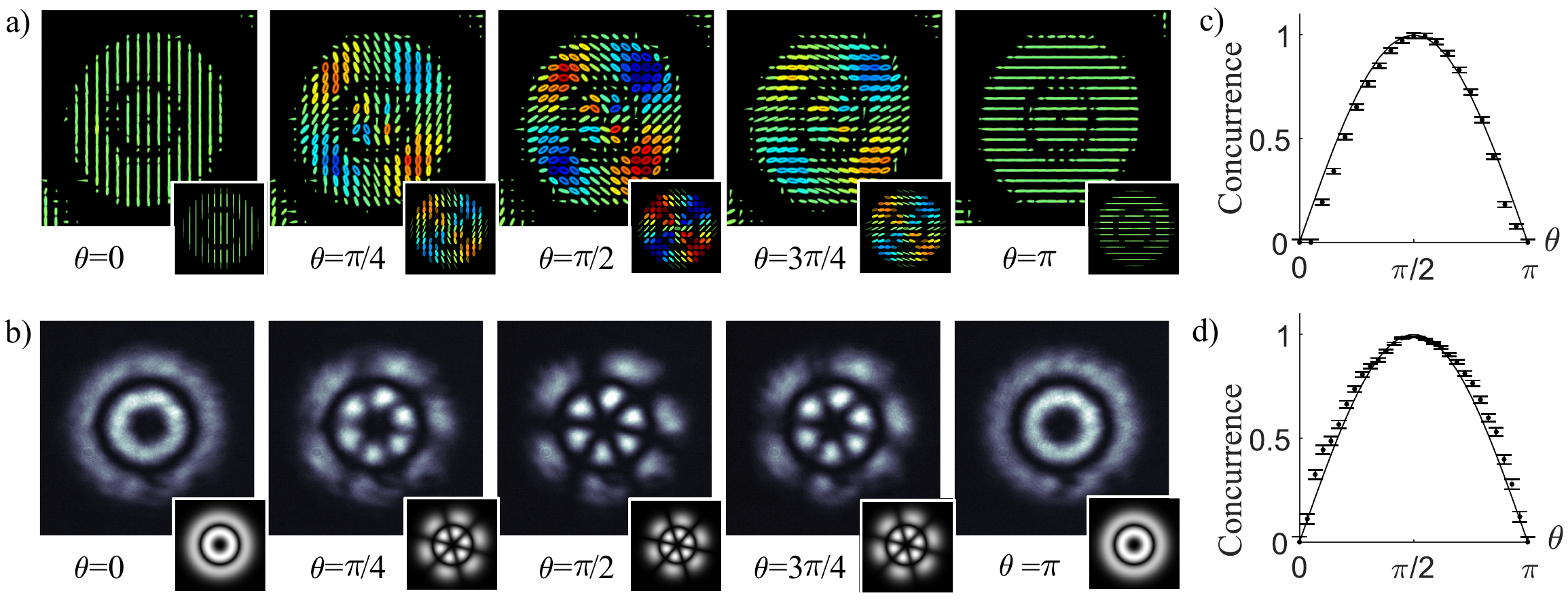}
\caption{Vector fields with varying degree of non-separability, characterized by concurrence measurement via basis-independent Stokes parameters (top row) or projection measurements (bottom row). 
Basis-independent concurrence measurements are shown for the example of a field $\cos (\theta/2) LG_{1}^{-1} \hat{v}+  \sin  (\theta/2) LG_{1}^{1} \hat{h}$, with examples of the measured and predicted (insets) polarization profiles shown in (a). The polarization ellipses are evaluated on a 21 by 21 grid, with red (blue) indicating right (left) circular polarization, and linear polarization shown in green. The associated concurrence is shown in (c). Projection measurements are shown in (b) for $ \cos (\theta/2) LG_{1}^{3} \hat{r}+  \sin  (\theta/2) LG_{1}^{-3} \hat{l}$, respectively, analyzed in terms of measured and predicted (insets) intensity profiles, and concurrence shown in (d). The calculations of error bars are discussed in the text.  \label{concurrence}}
\end{figure*}

We illustrate the potential of our technique by designing light beams with a varying degree of non-separability, based on the superposition of two orthogonal spatial modes, here illustrated for the case of different Laguerre-Gaussian modes, with radial and azimuthal mode numbers $p_{v,h}$ and $\ell_{v,h}$ respectively:  
\begin{align} \label{varying_vector_beam}
\vec{u}(\vec{r}_\perp) = \cos (\theta/2) LG_{pv}^{\ell v}(\vec{r}_\perp) \hat{v}+  \sin  (\theta/2) LG_{ph}^{\ell h}(\vec{r}_\perp) \hat{h}. 
\end{align}
\noindent
By varying $\theta$ in the interval $[0, \hspace{1mm} \pi]$ we can move from a uniformly vertically polarized beam at $\theta =0$ via a vector beam at $\theta=\pi/2$ to a uniformly horizontally polarized beam in the orthogonal polarization basis at $\theta=\pi$. 
Our spatially resolved Stokes parameter measurements clearly show this structure, as illustrated in Fig.~\ref{concurrence}a). We record the overall intensity in the 6 polarization states, integrated over the beam profile, to determine the concurrence according to (\ref{Concurrence_Experiment}), as shown in Fig.~\ref{concurrence}c). 


As mentioned above, detector noise tends to increase our estimate of concurrence. To combat this effect we introduce a low-pass spatial filter in our analysis, removing the two highest frequency components in the recorded images, and thus filtering out high spatial frequency noise while keeping the smoother varying actual spatial modes. We quantify the remaining uncertainty in our concurrence by recording the mean and standard deviation of 21 measurements.

We compare our results with concurrence measurements based on projective measurements according to a recently developed tomographic method \cite{Ndagano2016}.  In this case, the concurrence (or vector beam quality factor) can be deduced from the projection of the vector beam into 6 different spatial modes for two orthogonal polarizations each. The polarization states can be separated by polarizing optical elements, and the spatial states are determined by transforming the relevant spatial modes to a Gaussian mode via SLM and measuring on-axis intensity.  Small intensity fluctuations at the detector cause uncertainty in the measured concurrence which was characterized for each intensity measurement by taking the standard deviation after averaging over the 64 central pixels, as detailed in ref.~\cite{Ndagano2016}, leading to concurrence measurements shown in Fig.~\ref{concurrence}d).  Both methods are able to quantify the degree of non-separability at comparable accuracy.  

We have characterized the complex vectorial light fields using two complementary approaches:  a recently developed tomographic method based on projection measurements on the polarization and spatial degree of freedom, and the basis-independent measurement based on spatially resolved Stokes measurements introduced here.  The former approach requires at least twelve measurements, six of which are projections onto spatial modes which require custom optics and rely on prior knowledge of the spatial modes.  In contrast, our state-independent method characterizes the non-separability of two degrees of freedom by projective measurements on only one, the polarization. 
We note that while we have used a high resolution detector in order to reduce the systematic error introduced by detector noise, the experiments could have been equivalently performed by a single photodiode. The measurement speed in our setup is limited by the need to rotate waveplates, but this could be avoided by splitting the beam into at least 4 polarization components and simultaneously measuring them with 4 photo-detectors.

\textbf{Conclusion.}  Complex vectorial light fields are highly topical yet their measurement and characterization remains in its infancy.  Here we have outlined new approaches to the creation and detection of such light fields.  Using DMD technology we have demonstrated a fast, robust and inexpensive technique to create any vector field. The polarization insensitive DMDs are ideal candidates for the generation of complex light fields, a property that we have taken full advantage of here for the first time. 
We have demonstrated the ability to detect and characterize these fields using a basis-independent method which requires no prior knowledge of the incoming field. This represents a significant advance in the metrology of these complex fields, and addresses a previously unresolved issue, namely, that the concurrence should not change with a change of basis.  Our method uses conventional Stokes projections to infer a spatially resolved and global concurrence.  For quantum systems, concurrence is a widely used measure that characterizes the entanglement of bipartite mixed states, and that determines the suitability of states for certain quantum protocols.  One may expect that concurrence of classical vector beams similarly determines its use for applications that rely on non-separability between different degrees of freedom.

This work was supported by EPSRC (Award No. EP/M508056/1).

\bibliographystyle{osajnl}
\bibliography{References}
\end{document}